\documentstyle[twoside,fleqn,epsfig]{actaps}

\begin{document}
%%%%%%%%%%%%%%%%%%%%%%%%%%%%%%%%%%%%%%%%%%%%%%%%%%%%%%%%%%%%%%%%
%%%%%%%%%%%% personal definitions%%%%%%%%%%%%%%%%%%%%%%%%%%%%%%%
%%%%%%%%%%%%%%%%%%%%%%%%%%%%%%%%%%%%%%%%%%%%%%%%%%%%%%%%%%%%%%%%
 \newcommand{\beq}{\begin{equation}}\newcommand{\eeq}{\end{equation}}
 \newcommand{\barr}{\begin{eqnarray}}\newcommand{\earr}{\end{eqnarray}}
\newcommand{\andy}[1]{ }

 \def\tltl{\widetilde}
 \def\bmn{\mbox{\boldmath $n$}} \def\bmA{\mbox{\boldmath $A$}}
 \def\bmB{\mbox{\boldmath $B$}} \def\bmb{\mbox{\boldmath $b$}}
 \def\bmsigma{\mbox{\boldmath $\sigma$}}
 \def\bmsigman{\mbox{\boldmath $\sigma$}\cdot\mbox{\boldmath $n$}}
 \def\bmsigmab{\mbox{\boldmath $\sigma$}\cdot\mbox{\boldmath $b$}}
 \def\bmsigmaA{\mbox{\boldmath $\sigma$}\cdot\mbox{\boldmath $A$}}
 \def\bmsigmaB{\mbox{\boldmath $\sigma$}\cdot\mbox{\boldmath $B$}}
 \def\ch{\mbox{ch}}
 \def\sh{\mbox{sh}}
 \def\eqn#1{Eq.\ (\ref{eq:#1})}
 \def\coltwovector#1#2{\left({#1\atop#2}\right)}
 \def\up{\coltwovector10}
 \def\down{\coltwovector01}
 \newcommand{\bm}[1]{\mbox{\boldmath $#1$}}
 \newcommand{\bmsub}[1]{\mbox{\boldmath\scriptsize $#1$}}
 \newcommand{\ket}[1]{| #1 \rangle}
 \newcommand{\bra}[1]{\langle #1 |}
%%%%%%%%%%%%%%%%%%%%%%%%%%%%%%%%%%%%%%%%%%%%%%%%%%%%%%%%%%%%%%%
%%%%%%%%%%%%%%%%%%%%%%%%%%%%%%%%%%%%%%%%%%%%%%%%%%%%%%%%%%%%%%%

% three definitions for page headings
\headings{1}{8}
\def\authorlist{P.\ Facchi, S.\ Pascazio}
\def\shorttitle{Berry phase due to quantum measurements}

\title{\uppercase{Berry phase due to quantum measurements}}

\author{P.\ Facchi\email{paolo.facchi@ba.infn.it},
S.\ Pascazio\email{saverio.pascazio@ba.infn.it}}
{
Dipartimento di Fisica, Universit\`a di Bari \\
and Istituto Nazionale di Fisica Nucleare, Sezione di Bari \\
 I-70126  Bari, Italy
}

\day{30 April 1999}  %Submission date

\abstract{%
The usual, ``static" version of the quantum Zeno effect consists in the
hindrance of the evolution of a quantum systems due to repeated
measurements. There is however a ``dynamic" version of the same
phenomenon, first discussed by von Neumann in 1932 and subsequently
explored by Aharonov and Anandan, in which a system is forced to follow
a given  trajectory. A Berry phase appears if such a trajectory is a
closed loop in the projective Hilbert space.
A specific example involving neutron spin is considered
 and a similar situation with photon polarization is investigated.
}

\medskip


 \setcounter{equation}{0}
 \section{Introduction }
 \label{sec-introd}
 \andy{intro}

The usual, {\em static} version of the quantum Zeno effect (QZE) consists in
hindering (and eventually halting) the time evolution of a quantum system by
repeatedly checking if it has decayed \cite{von}.
In a few words, this is due to the fact that in time $dt$, by the Schr\"odinger equation,
the phase of a state $\psi(t)$ changes by $\hbox{O}(dt)$ while the
absolute value of its scalar product with the initial state changes by
$\hbox{O}(dt^2)$.
The {\it dynamic\/} quantum Zeno effect exploits the above features and
forces the evolution through an arbitrary trajectory by a series of repeated
measurements \cite{von1,AA87}: Let there be a family of states
$\phi_k$, $k=0,1,\ldots, N$, such that $\phi_0=\psi(0)$, and such that
successive states differ little from one another (i.e.,
$|\langle\phi_{k+1} | \phi_k \rangle|$ is nearly 1). Now let $\delta T =
T/N$ and at $T_k=k\delta T$ project the evolving wave function on
$\phi_k$. Then for sufficiently large $N$, $\psi(T) \approx \phi_{_N}$.
[The static QZE is the special case $\phi_k=\phi_0 (=\psi(0)) \ \forall \
k$.]

 In the following we will show how guiding a
system through a closed loop in its state space (projective Hilbert
space) leads to a geometrical phase
\cite{AA87,Panchar,BerryQuantal}. We will first summarize some
results valid for neutron spin \cite{continous,Berry} and then
consider the case of photon polarization \cite{BerryKlein}.

\section{Neutron spin}
\label{sec-neutron}
\andy{neutron}

 Assume first that there is {\em no} Hamiltonian acting on the system: the
 neutron crosses a region where no magnetic
field is present. It starts with spin up along the $z$-axis and is
projected on the family of states
 \andy{projfamily}
 \beq
 \phi_k \equiv \exp(-i\theta_k\bmsigman)\up  \qquad
   \hbox{with~} \theta_k \equiv \frac{ak}N  \;,
  \qquad k=0,\ldots,N \ ,
 \label{eq:projfamily}
 \eeq
 where $\bmsigma$ is the vector of the Pauli matrices and $\bmn =
(n_x,n_y,n_z)$ a unit vector (independent of $k$).

The neutron evolves for a time $T$ with projections at times $T_k =
k\delta T$ ($k=1,\dots,N$ and $\delta T=T/N$). The final state is
$\left[\phi_0 = \up\right]$
 \andy{finstate}
 \barr
 \ket{\psi(T)}
  &=& |\phi_N\rangle \langle \phi_N|
       \phi_{N-1}\rangle \cdots \langle \phi_2|
       \phi_1\rangle \langle \phi_1| \phi_0\rangle \nonumber \\
%  &=& |\phi_N\rangle
%       \left(\cos \frac{a}{N} + i n_z \sin \frac {a}{N} \right)^N \nonumber \\
  &=& \cos^N \left(\frac{a}{N} \right)
       \left(1 + i n_z \tan \frac {a}{N} \right)^N
       |\phi_N\rangle  \nonumber \\
  &\stackrel{N\rightarrow \infty}{\longrightarrow} &
%  \exp (ia n_z)
%     |\phi_N\rangle  \nonumber \\
%         & = &
  \exp (ia n_z) \exp (-ia\bmsigman) | \phi_0\rangle .
 \label{eq:finstate}
 \earr
 If
$a=\pi$, \andy{finstatepi}
 \beq
 \psi(T)
% = \exp (i \pi \cos \Theta) (-1) \phi_0
        = \exp [-i \pi (1-\cos \Theta)]  \phi_0
        = \exp (-i \Omega/2 )  \up ,
 \label{eq:finstatepi}
 \eeq
 where $\cos\Theta \equiv n_z$ and $\Omega$ is the solid angle
 subtended by the curve traced by the
spin during its evolution. The factor $ \exp (-i\Omega/2)$ is a Berry
phase and it is due only to measurements (the Hamiltonian is zero).
Notice that, as discussed by Pati and Lawande \cite{Pati}, no Berry phase appears
in the usual quantum Zeno context,
namely when $\phi_k \propto \phi_0 \ \forall \ k$, because in that case
$a=0$ in (\ref{eq:finstate}).

We now look at the process (\ref{eq:finstate}) for $N$
finite. The spin goes back to its initial state after describing a
regular polygon on the Poincar\'e sphere, as in Figure 1a.
 \begin{figure}
\centerline{\epsfig{file=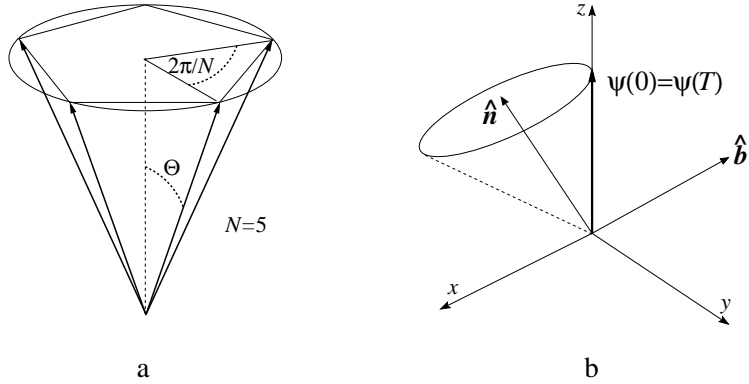, height=5cm}}
%\begin{center}
 \caption{Fig. 1. a) Spin evolution due to $N=5$ measurements.
b) Spin evolution with very frequent measurements and non-zero Hamiltonian.}
%\end{center}
 \end{figure}
 After $N (<\infty)$ projections the final state is \cite{Berry}
 \beq
 \psi(T)=\rho_N \exp(-i\beta_N)\phi_0,
 \eeq
 where
 \andy{rhoN, betaN}
 \barr
 \rho_N = \left(\cos^2\frac{\pi}{N}+n^2_z
 \sin^2\frac{\pi}{N}\right)^{\frac{N}{2}}, \qquad
% \label{eq:rhoN}\\
 \beta_N = \pi-N\arctan\left(\cos\Theta\tan\frac{\pi}{N}\right).
 \label{eq:betaN}
 \earr
The quantity $\rho_N$ accounts for the probability loss ($N$ is
finite and there is no QZE). It is easy to check that in the
``continuous measurement" limit (QZE) we recover the result
(\ref{eq:finstatepi}).

 The relation between the solid angle and the geometrical phase is valid
also with a finite number of polarizers $N$. Indeed, it is
straightforward to show that the solid angle subtended by  a regular $N$-sided polygon
(Figure 1a) is
 \beq
 \Omega_{N}
 %=N\Omega_{2\pi/N}
 =2\pi-2N\arctan\left(\cos\Theta\tan\frac{\pi}{N}\right)=2\beta_N.
 \eeq
This result is of course in agreement with other analyses \cite{SM}
based on the Pancharatnam connection \cite{Panchar}.

 Let us now consider the effect of a non-zero Hamiltonian
 (neutron spin in a magnetic field)
 \andy{Hamadd}
 \beq
H=\mu \bmsigmab ,
 \label{eq:Hamadd}
 \eeq
 where $\bmb = (b_x,b_y,b_z)$ is a unit vector, in general different from
$\bmn$. See Figure 1b.

If the system starts with spin up it has the
following ``undisturbed" evolution
 \andy{undisturb}
 \beq
 \psi(t) = \exp(-i\mu t\bmsigmab)\phi_0 .
 \label{eq:undisturb}
 \eeq
  Now let the system evolve for a time $T$ with projections at times
$T_k=k\delta T$ ($k=1,\dots,N$ and $\delta T=T/N$) and Hamiltonian
evolution in between. It is  not difficult to show that, in the
continuum limit ($N\to\infty$), the final state reads \cite{Berry}:
 \andy{finpsi}
 \barr
 \psi(T)
=\exp\left(-i \int_0^T \langle \psi(t) | H | \psi(t) \rangle dt\right)
\exp\left(i a n_z  - ia\bmsigman \right) \phi_0.
 \label{eq:finpsi}
 \earr
 The first factor in (\ref{eq:finpsi}) is obviously the dynamical
phase and the remaining phase, when the spin goes back to its
initial state, is the geometrical phase: when $a=\pi$
 \andy{fun}
 \beq
 \psi(T) = \exp \left( - i\Omega/2 \right)
 \exp\left(- i\mu T (\bmb \cdot \bmn) n_z \right) \up ,
 \label{eq:fun}
 \eeq
 where $\Omega$ is the solid angle subtended by the curve traced out by
the spin, as in (\ref{eq:finstatepi}), and $\mu T (\bmb \cdot \bmn)
n_z$ is the dynamical phase.

\section{Photon polarization}
\label{sec-photon}
\andy{photon}
The experiments described in Section \ref{sec-neutron} involve
neutrons, or in general {\it massive} spin-$1/2$ particles. This
allowed us to neglect the neutron momentum in our analysis: the
axis of the neutron polarizer can have an arbitrary direction with
respect to the neutron momentum. Equations very similar to those of
Section \ref{sec-neutron} were obtained by Berry and Klein in their
beautiful work on polarized light \cite{BerryKlein}. However, when
one deals with photons, the additional constraint of transversality
makes things more complicated: the photon polarization must be
perpendicular to momentum and for this reason, in Ref.\
\cite{BerryKlein}, the projection on states of non-linear
polarization is achieved by making use of ``retarders" \cite{SiMu}.

It is interesting, in this context, to discuss a geometric
configuration proposed by A.G.\ Klein \cite{Klein1}. A photon is
sent into a polygonal cylinder made up of $N$ perfect plane mirrors
and emerges after $N$ reflections: see Figure 2 ($N=6$).
 \begin{figure}
\centerline{\epsfig{file=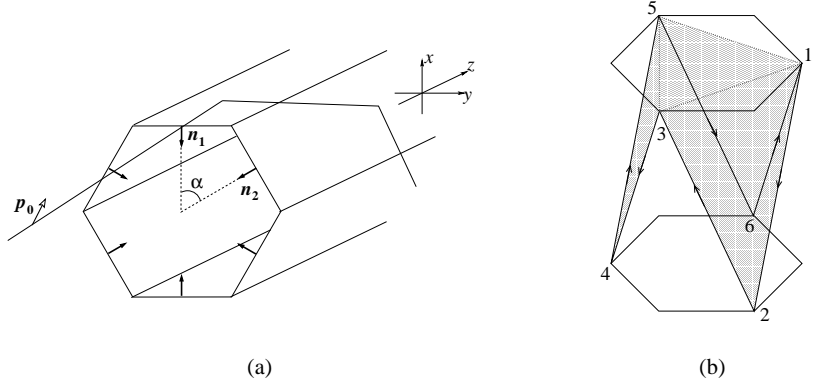, height=5cm}}
%\vspace{5cm}
%\begin{center}
 \caption{Fig. 2. (a) Klein's arrangement: a photon is sent into a polygonal
cylinder, bouncing off $N=6$ perfect mirrors and emerging from the
opposite side. (b) Evolution of the polarization. The points 1-6
represent the tip of the polarization vector and the shadow area
yields the Berry phase. }
%\end{center}
 \end{figure}
Let us consider a photon with momentum parallel to $\hat{\bm
k}=(\sin\theta\cos\phi,\sin\theta\sin\phi,\cos\theta)$ and
polarization $\ket{p}$, which is reflected by an ideal mirror with
normal unit vector $\bm n$. After reflection, the photon helicity
is reversed and the momentum becomes
\beq
\hat{\bm k}'=-R_{\bmsub n}(\pi)\hat{\bm k},
\eeq
where $R_{\bmsub n}(\phi)$ represents a rotation by an angle $\phi$
around direction $\bm n$. Hence, after a reflection, the
polarization state becomes (ideal mirror, infinite conductivity)
\andy{reflection}
\beq\label{eq:reflection}
\ket{p'}=M(\bm n)\ket{p},\quad\mbox{with}\quad
M(\bm n)=\exp\left(-i\bm n\cdot\bm J\pi\right),
\eeq
where $\bm J=(J_x,J_y,J_z)$ are the generators of rotations in {\bf
R}$^3$. Consider now a polygonal cylinder of ideal mirrors, whose
normal vectors are
\andy{mirror}
\beq\label{eq:mirror}
\bm n_1=
  \left(-1,0,0\right),
\quad \bm n_\ell= R_z[(\ell-1)\alpha]\bm n_1
=[-\cos(\ell-1)\alpha,-\sin(\ell-1)\alpha,0],
\eeq
where $\alpha=2\pi/N$ for a regular $N$-sided polygon [see
Figure~2(a)]. The operator representing the action of the $\ell$-th
mirror is $M_\ell=M(\bm n_\ell)$. By using (\ref{eq:reflection})
and (\ref{eq:mirror}) one gets
%[$M_1=-\sigma_x$]
\beq
M_\ell
%=e^{-i\alpha(\ell-1) \sigma_z/2} M_1 e^{i\alpha(\ell-1) \sigma_z/2}
=-\exp\left(-i\alpha(\ell-1) J_z\right)
M_1 \exp\left(i\alpha(\ell-1) J_z\right),\quad M_1=\exp(i\pi J_x).
\eeq
Let $\ket{p_0}$
be the initial photon polarization; after $N$
reflections, the photon emerges with final polarization
\andy{finpol}
\beq\label{eq:finpol}
\ket{p_N}=M_N\cdots M_2 M_1 \ket{p_0}
=\exp \left( -i N \alpha J_z \right)\tltl M^N\ket{p_0},
\eeq
where
\beq
\tltl M=\exp\left(i\alpha J_z\right) M_1
= \exp\left(i\alpha J_z\right) \exp(i\pi J_x).
\eeq
By using $[J_i, J_j]=i\varepsilon_{ijk}J_k$, we obtain
\andy{nrot}
\beq\label{eq:nrot}
\tltl M=\exp\left(-i\pi\tilde{\bm n}\cdot\bm J\right),
\quad\mbox{with}\quad
\tilde{\bm n}=\left(-\cos\frac{\alpha}{2},\sin\frac{\alpha}{2},0 \right),
\eeq
so that $\tltl M$ is a $\pi$ rotation around
$\tilde{\bm n}$. Remembering that $N\alpha=2\pi$ and using
(\ref{eq:nrot}), the final polarization (\ref{eq:finpol}) reads
\beq
\ket{p_N}=\exp(-i 2\pi J_z)\exp\left(-i N\pi
\tilde{\bm n}\cdot\bm J \right)\ket{p_0}
=  \exp\left(-i N\pi
\tilde{\bm n}\cdot\bm J \right)\ket{p_0} .
\eeq
For odd $N=2m+1$, one gets
\beq
\ket{p_N}
=\exp\left(-i (2m+1)\pi \tilde{\bm n}\cdot\bm J\right)\ket{p_0}
=\exp\left(-i \pi
\tilde{\bm n}\cdot\bm J\right)\ket{p_0}
\eeq
and the final polarization is not in the same direction as the
initial one. On the other hand, for even $N=2m$, the polarization
vector does describe a closed loop, but one obtains
\beq
\ket{p_N}=\exp(-i 2m \pi \tilde{\bm n}\cdot\bm J)\ket{p_0}
=\ket{p_0}
\eeq
and the photon acquires no geometrical phase. The reason for this
result is shown in Figure~2(b): for even $N$, the polarization
vector {\it always} describes a solid angle $\Omega=2\pi$: the
photon always acquires a Berry phase $\beta=\Omega=2\pi$, with no
physical effects.

The experiment just described is not equivalent to the one analyzed
in the previous section. Indeed, the dynamics of reflections is
always unitary. The difficulty in obtaining a geometrical phase is
due to the condition of transversality of the electromagnetic
field. To encompass this situation one needs (at least three)
mirrors whose normal vectors do not lie in the same plane, as shown
in \cite{Kitano}.

\medskip
\noindent {\bf Acknowledgments:}
We thank A.G.\ Klein and L.S.\ Schulman for many useful discussions
and M.V.\ Berry and A.K.\ Pati for interesting comments on Ref.\
\cite{Berry}. The mirror geometry investigated in
Section~\ref{sec-photon} was proposed by A.G.\ Klein.

%%%%%%%%%%%%%%%%%%%%%%% REFERENCES %%%%%%%%%%%%%%%%%%%%%%%%%%%%%%%

 \end{document}